\documentclass[preprint2]{aastex}

\usepackage{xspace}

\newcommand{\ye}{{\ensuremath{Y_{\mathrm{e}}}}\xspace}
\newcommand{\g}{{\ensuremath{\mathrm{g}}}\xspace}

\newcommand{\cm}{{\ensuremath{\mathrm{cm}}}\xspace}

\newcommand{\km}{{\ensuremath{\mathrm{km}}}\xspace}
\newcommand{\Msun}{{\ensuremath{\mathrm{M}_{\odot}}}\xspace}

\newcommand{\gcc}{{\ensuremath{\g\,\cm^{-3}}}\xspace}

\newcommand{\lSect}[1]{{\label{sec:#1}}}

\newcommand{\ltaprx} {\lower .1ex\hbox{\rlap{\raise .6ex\hbox{\hskip .3ex
        {\ifmmode{\scriptscriptstyle <}\else 
                {$\scriptscriptstyle <$}\fi}}}
        \kern -.4ex{\ifmmode{\scriptscriptstyle \sim}\else 
                {$\scriptscriptstyle\sim$}\fi}}}
\newcommand{\gtaprx} {\lower .1ex\hbox{\rlap{\raise .6ex\hbox{\hskip .3ex
        {\ifmmode{\scriptscriptstyle >}\else 
                {$\scriptscriptstyle >$}\fi}}}
        \kern -.4ex{\ifmmode{\scriptscriptstyle \sim}\else 
                {$\scriptscriptstyle\sim$}\fi}}}

\slugcomment{Version \today}

\shorttitle{Nucleosynthesis in the Hot Bubble}
\shortauthors{Pruet, Woosley, \& Janka}

\begin{document}

\title{Nucleosynthesis in the Hot Convective Bubble in Core-Collapse
Supernovae}

\author{J. Pruet}
\affil{N Division, Lawrence Livermore National Laboratory, P. O.Box
  808, Livermore, CA 94550}
\email{pruet1@llnl.gov}
\author{S. E. Woosley}
\affil{Department of Astronomy and Astrophysics, UCSC, Santa Cruz, CA,
95064}
\email{woosley@ucolick.org}
\author{R. Buras}
\affil{Max-Planck-Institut f\"ur Astrophysik, Karl-Schwarzschild-Str. 1,
85741 Garching, Germany}
\email{rburas@mpa-garching.mpg.de}
\author{H.-T. Janka}
\affil{Max-Planck-Institut f\"ur Astrophysik, Karl-Schwarzschild-Str. 1,
85741 Garching, Germany}
\email{thj@mpa-garching.mpg.de}
\and
\author{R.D. Hoffman}
\affil{N Division, Lawrence Livermore National Laboratory, P. O.Box
  808, Livermore, CA 94550}
\email{rdhoffman@llnl.gov}

\begin{abstract}
As an explosion develops in the collapsed core of a massive star,
neutrino emission drives convection in a hot bubble of radiation,
nucleons, and pairs just outside a proto-neutron star. Shortly
thereafter, neutrinos drive a wind-like outflow from the neutron star.
In both the convective bubble and the early wind, weak interactions
temporarily cause a proton excess ($\ye \gtaprx 0.50$) to
develop in the ejected matter. This situation lasts for at least the
first second, and the
approximately 0.05 - 0.1 $\Msun$ that is ejected has an unusual
composition that may be important for nucleosynthesis. Using tracer
particles to follow the conditions in a two-dimensional model of a
successful supernova explosion calculated by \citet{jan03}, we
determine the composition of this material.  Most of it is helium and
$^{56}$Ni. The rest is relatively rare species produced by the decay
of proton-rich isotopes unstable to positron emission. In the absence
of pronounced charged-current neutrino capture, nuclear flow will
be held up by long-lived waiting point nuclei in the vicinity of
$^{64}{\rm Ge}$. The resulting abundance pattern can be modestly rich
in a few interesting rare isotopes like $^{45}{\rm Sc}$, $^{49}{\rm
Ti}$, and $^{64}{\rm Zn}$. The present calculations imply yields that,
when compared with the production of major species in the rest of the
supernova, are about those needed to account for the solar abundance
of $^{45}{\rm Sc}$ and $^{49}{\rm Ti}$. Since the synthesis will be
nearly the same in stars of high and low metallicity, the primary
production of these species may have discernible signatures in the
abundances of low metallicity stars.  We also discuss uncertainties in
the nuclear physics and early supernova evolution to which abundances
of interesting nuclei are sensitive.
\end{abstract}


\section{INTRODUCTION}

When the iron core of a massive star collapses to a neutron star, a
hot proto-neutron star is formed which radiates away its final binding
energy as neutrinos. Interaction of these neutrinos with the infalling
matter has long been thought to be the mechanism responsible for
exploding that part of the progenitor external to the neutron star and
making a supernova (e.g., Janka 2001; Woosley, Heger, \& Weaver 2002, 
and references therein).
During the few tenths of a
second when the explosion is developing, a convective bubble of
photo-disintegrated matter (nucleons), radiation, and pairs lies above
the neutron star but beneath an accretion shock. Neutrino interactions
in this bubble power its expansion, drive convective overturn, and determine
its composition. Since baryons exist in the bubble only as nucleons,
the critical quantity for nucleosynthesis is the proton mass fraction
($\ye$). Initially, in part because of an excess of electron neutrinos
over antineutrinos, $\ye \gtaprx 0.5$ \citep{qia96}. As time passes,
however, the fluxes of the different neutrino flavors and their
spectra change so that $\ye$ evolves and becomes considerably less
than 0.5. This epoch, also known as the ``neutrino-powered wind'', has
been explored extensively as a possible site for the r-process
\citep{qia96,hof97,woo94,car97,qia00, tak94,ots00,sum00,Tho01} as well
as $^{64}$Zn and some light p-process nuclei \citep{hof96}.

In this paper we consider nucleosynthesis during the earlier epoch
when $\ye$ is still greater than 0.5. This results in a novel
situation in which the alpha-rich freeze out occurs in the presence of
a non-trivial abundance of free protons. The resulting nuclear flows
thus have characteristics of both the alpha-rich freeze out
\citep{Woo73,Woo92} and the rp-process \citep{Wal81}. Several
proton-rich nuclei, e.g., $^{64}$Ge and $^{45}$Cr, are produced in
such great abundance that, after ejection and decay, they contribute a
significant fraction of the solar inventory of such species.

\section{Supernova Model and Nuclear Physics Employed}

\subsection{Explosion Model for a 15$\,$M$_{\odot}$ Star}

The nucleosynthesis calculations in this paper are based on
a simulation of the neutrino-driven explosion
of a nonrotating 15$\Msun$ star (Model S15A of Woosley \& Weaver
1995) by \citet{jan03} (see also Janka et al.\ 2004).
The post-bounce evolution of the model was
followed in two dimensions (2D) with a polar coordinate grid
of 400 (nonequidistant) radial zones and 32 lateral zones
(2.7 degrees resolution),
assuming azimuthal symmetry and using periodic conditions at
the boundaries of the lateral wedge at $\pm 43.2^{\mathrm{o}}$
above and below the equatorial plane. Convection was seeded in
this simulation by velocity perturbations of order $10^{-3}$,
imposed randomly on the spherical post-bounce core.

The neutrino transport was decribed by solving the
energy-dependent neutrino
number, energy, and momentum equations in radial direction in
all angular bins of the grid, using closure relations from a
model Boltzmann equation \citep{ram02}.
Neutrino pressure gradients and neutrino advection in lateral
direction were taken into account (for details, see 
Buras et al.\ 2004).
General relativistic effects were approximately included as
described by \citet{ram02}. 

Although convective activity develops
in the neutrino-heating layer behind the supernova (SN) shock
on a time scale of several ten milliseconds after bounce, no
explosions were obtained with the described setup until
$\sim$250$\,$ms \citep{bur03}, at which time the very CPU-intense 
simulations usually had to be terminated. 
The explosion in the simulation
discussed here was a consequence of omitting the velocity-dependent
terms from the neutrino momentum equation. This manipulation 
increased the neutrino-energy density und thus the neutrino 
energy deposition in the heating region by $\sim$20--30\%
and was sufficient to convert a failed model into an 
exploding one (see also Janka et al.\ 2004, Buras et al.\ 2004). 
This sensitivity of the outcome of the 
simulation to only modest changes of the transport treatment 
demonstrates how close the convecting, 2D models of \citet{bur03}
with energy-dependent neutrino transport are to ultimate success.

\begin{figure}
\epsscale{1.5}
\plotone{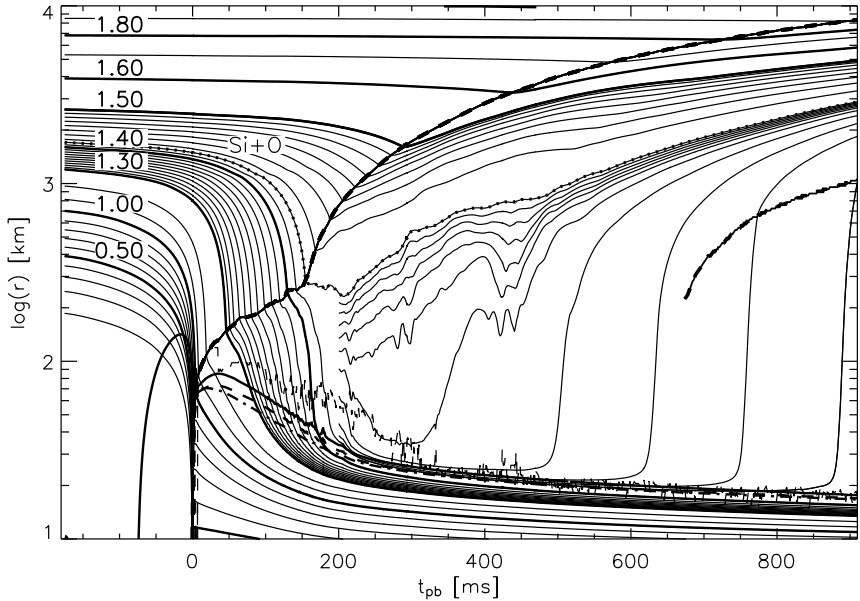}
\caption{Radius versus post-bounce time for selected ``mass shells'' in the
exploding SN model of \citet{jan03}, spaced with intervals between
0.001$\,$M$_{\odot}$ and 0.1$\,$M$_{\odot}$.
The simulation was carried out
in two dimensions until about 470$\,$ms after bounce, and was continued
in spherical symmetry thereafter. During the 2D period the lines
do not trace the trajectories of Lagrangian mass elements but
correspond to the radii of spheres enclosing certain values of 
the integrated rest mass. The expanding
dashed lines mark the positions of forward shock and wind termination
shock, respectively, and the thin dashed, wiggled line represents the
angularly averaged position of the gain radius. The
neutrinosphere positions of electron neutrinos (solid), electron
antineutrinos (dashed), and muon and tau neutrinos and antineutrinos
(dash-dotted) are also indicated.
The explosion sets in when the shock passes the
infalling interface between the Si layer and the oxygen-enriched Si
layer (given by the trajectory marked with dots)
at which the density begins to drop steeply and the entropy increases
from about 2.5 to nearly 5$\,k_b$ per nucleon.   
\label{massshells}}
\end{figure}

\begin{figure}
\epsscale{1.5}
\plotone{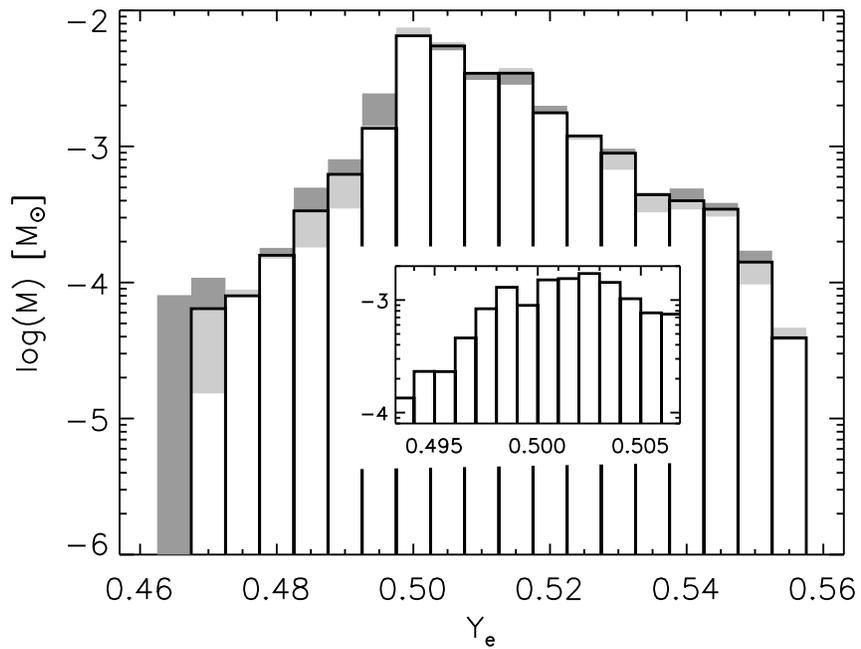}
\caption{Ejecta mass versus $\ye$ of neutrino-heated and
-processed matter during the convective phase until $\sim$470$\,$ms  
post bounce. The insert shows the
region around $\ye\sim 0.5$ in higher resolution. The
grey shading indicates estimated errors due to the limited spatial
resolution of the two-dimensional simulation (for details, see
Buras et al.\ 2004).
\label{massye}}
\end{figure}

The evolution from the onset of core collapse (at about $-175\,$ms)
through core bounce and convective phase to
explosion is shown in terms of mass shell trajectories in 
Fig.~\ref{massshells}. The explosion sets in when the  
infalling interface between Si layer and oxygen-enriched 
Si layer reaches the shock at about 160$\,$ms post bounce.
The corresponding steep drop of the density and mass accretion
rate, associated with an entropy increase by a factor of $\sim\,$2,
allow the shock to expand and convection to become more 
violent, thus establishing runaway conditions. The calculation
was performed in 2D for following the ejection of the 
convective shell until 470$\,$ms after bounce. While matter
is channeled in narrow downflows towards the gain radius,
where it is heated by neutrinos and some of it starts 
expanding again in high-entropy bubbles, its neutron-to-proton
ratio is set by weak interactions with electron neutrinos and
antineutrinos as well as electron and positron captures on free
nucleons. The final value of $\ye$ is a crucial 
parameter for the subsequent nucleosynthesis. The mass distribution
of neutrino-heated and -processed ejecta from the convective bubble
is plotted in Fig.~\ref{massye}.

\begin{figure*}
\epsscale{1.0}
\plotone{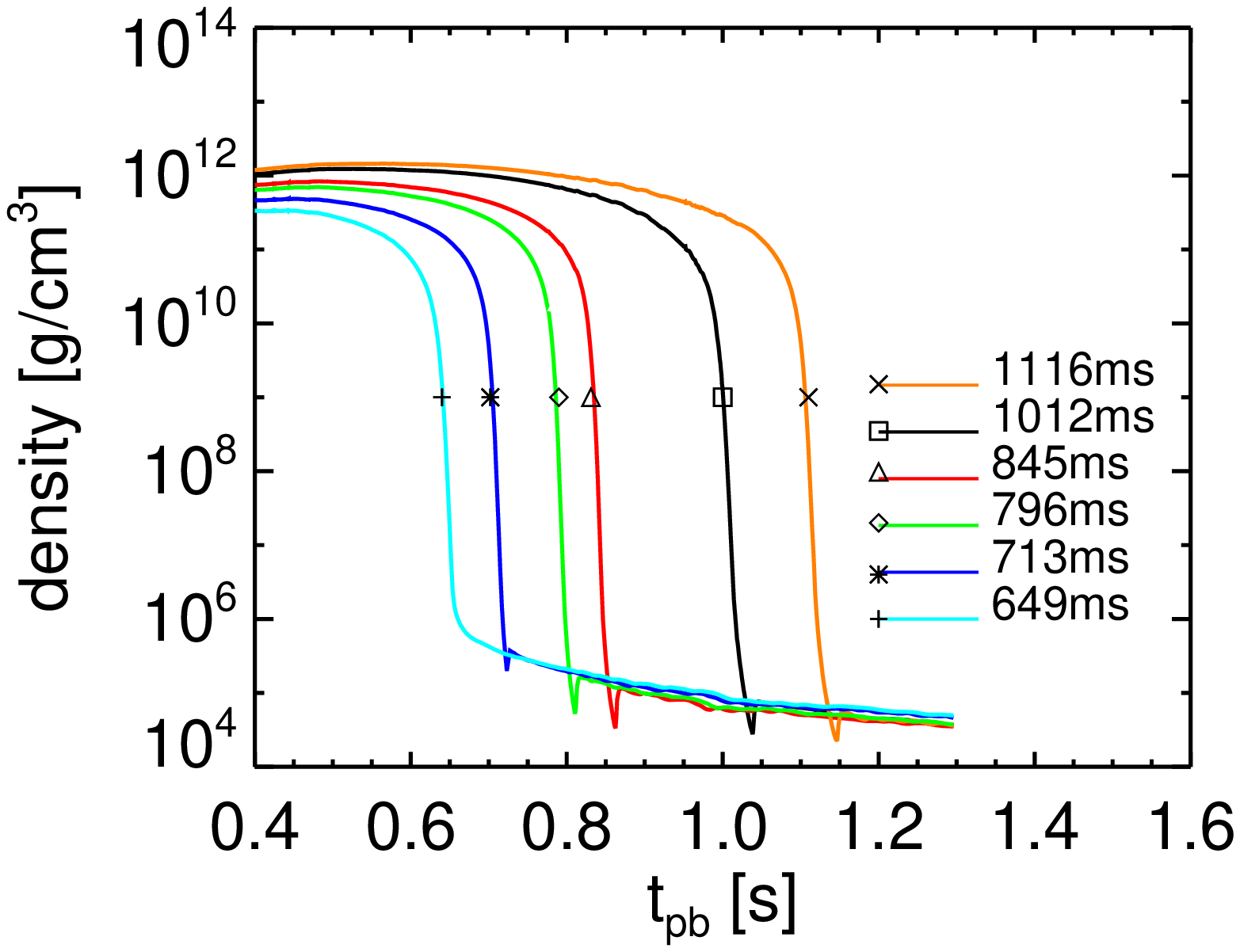}
\plotone{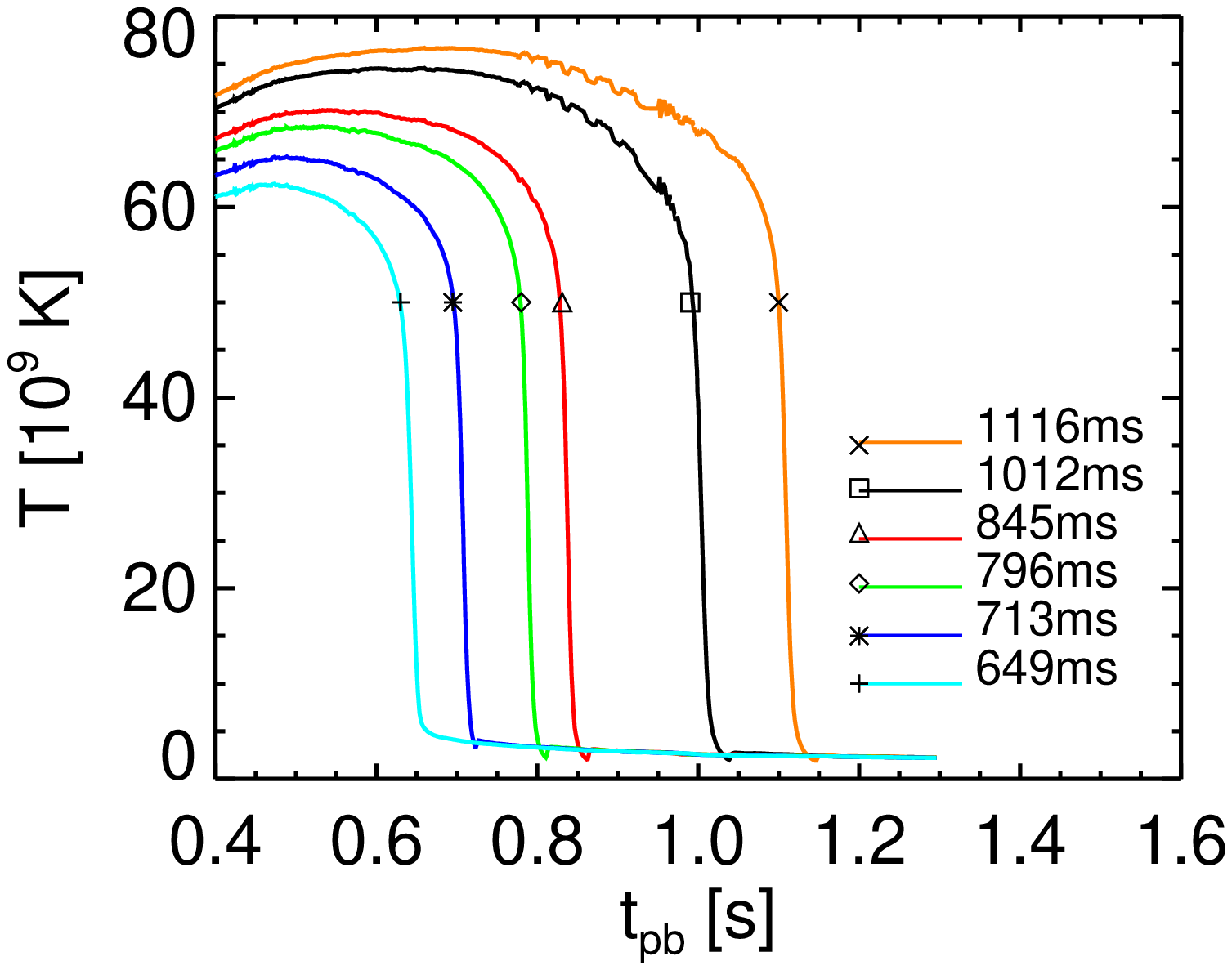}\\
\plotone{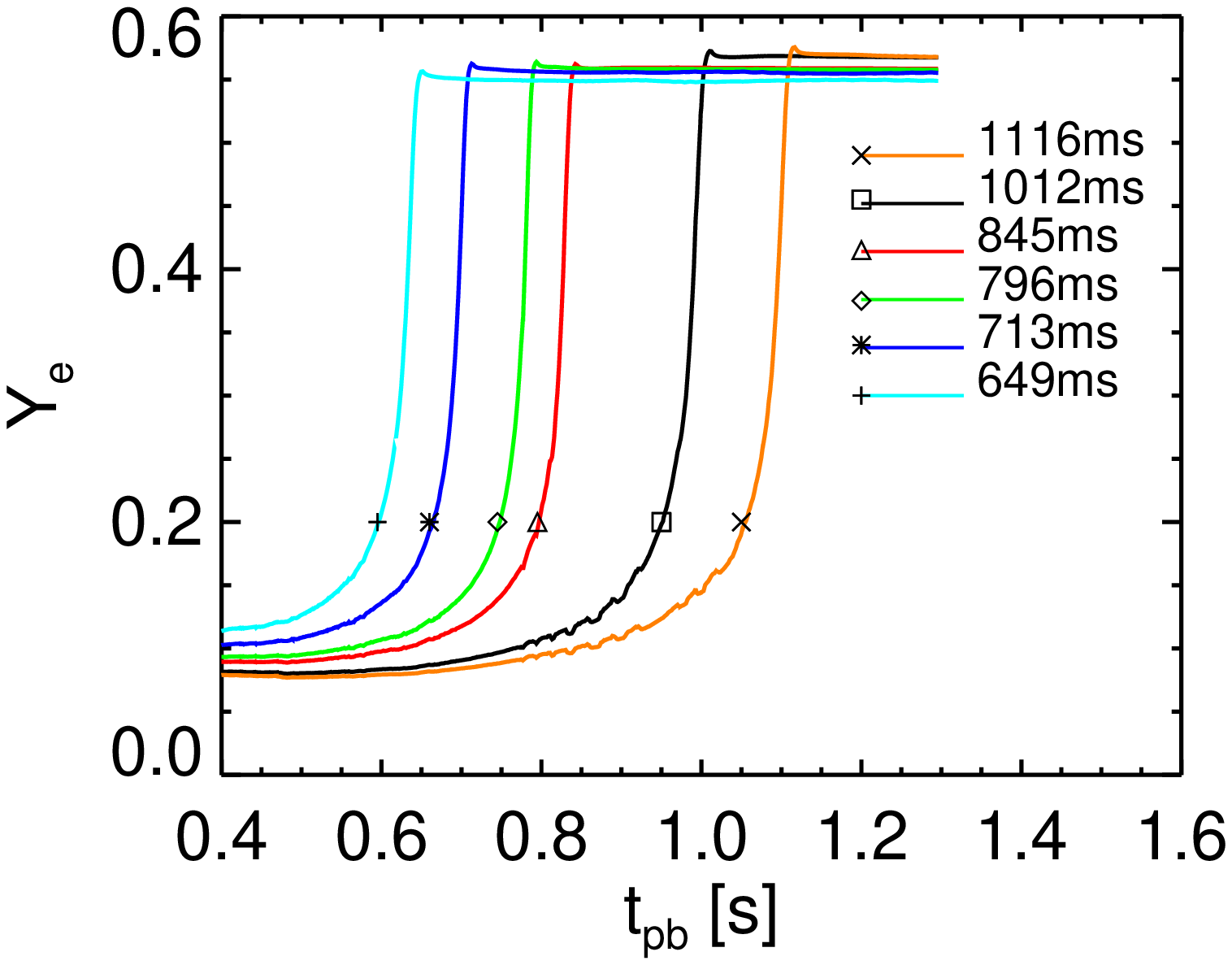}
\plotone{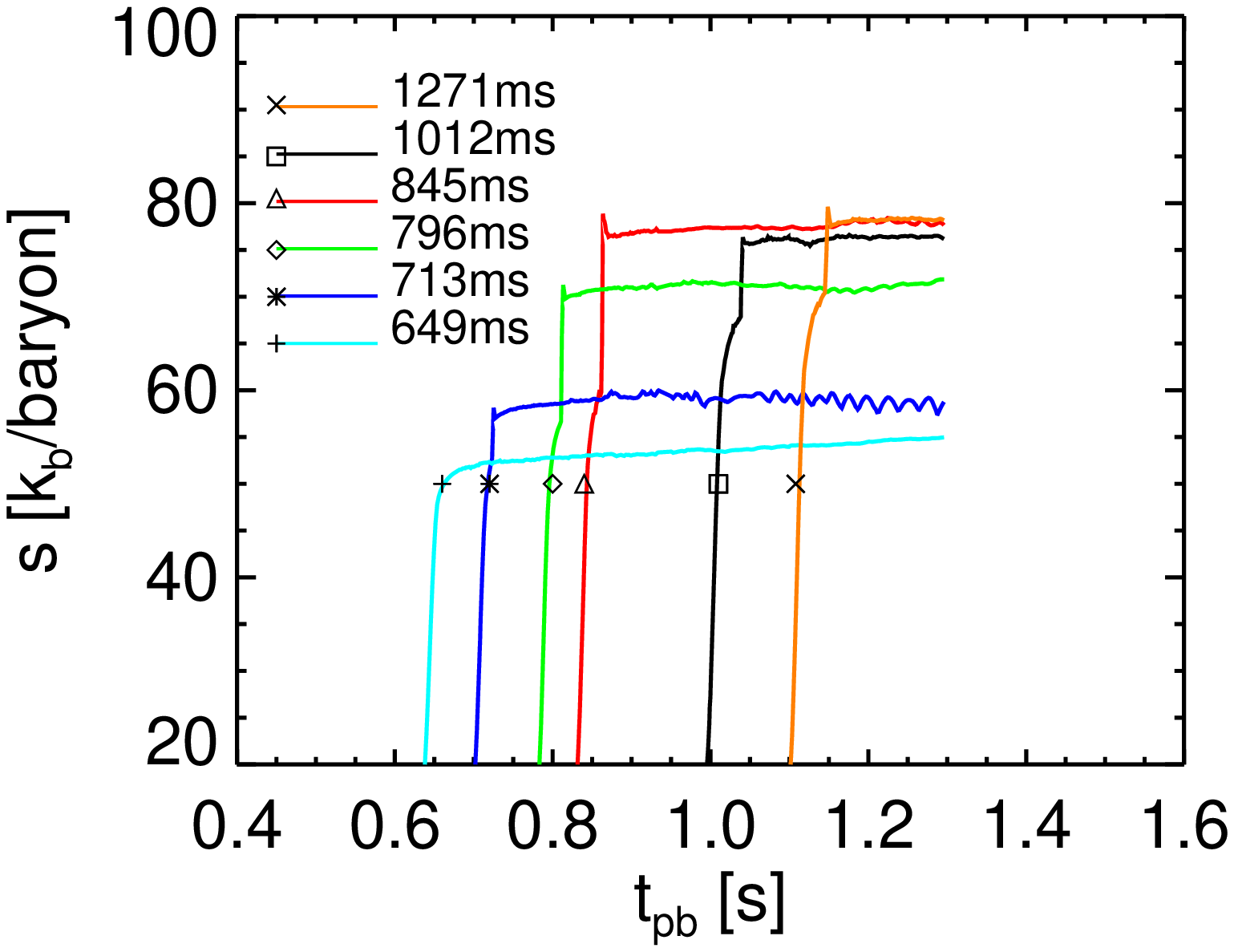}
\caption{Density, temperature, $\ye$, and entropy as
functions of post-bounce time along the trajectories of mass elements
around an enclosed baryonic mass of 1.41$\,$M$_{\odot}$. The elements
first follow the rise of  temperature and density in the outer layers
of the contracting neutron star and then enter a phase of very rapid
expansion when they are ejected in the neutrino-driven wind. The
curves are labeled by the time the mass elements
cross a radius of 100$\,$km. The collision with the slower preceding
ejecta occurs through a wind termination
shock and is visible as a non-monotonicity of the density and temperature,
associated with an entropy increase of 10--15$\,k_b$ per nucleon.
\label{wind}}
\end{figure*}

At 470$\,$ms after bounce the model
was mapped to a 1D grid and the subsequent evolution 
was simulated until 1300$\,$ms after bounce.
With accretion flows to the neutron star having ceased, this
phase is characterized by an essentially spherically symmetric
outflow of matter from the nascent neutron star, which is driven 
by neutrino-energy deposition outside the neutrinosphere
\citep{woo92,dun86}. This neutrino-powered wind is visible 
in Fig.~\ref{massshells} after $\sim$500$\,$ms. The fast
wind collides with the dense shell of slower ejecta behind the
shock and is decelerated again. The corresponding negative 
velocity gradient steepens to a reverse shock when the wind
expansion becomes supersonic (Fig.~\ref{massshells}; 
Janka \& M\"uller 1995). Characteristic parameters for some
mass shells in this early wind phase are shown in Fig.~\ref{wind}. 
Six representative shells are sufficient, because
the differences between the shells evolve
slowly with time according
to the slow variation of the conditions (neutron star
radius, gravitational potential, neutrino luminosities and 
spectra) in the driving region of the wind near the neutron 
star surface. In Table~\ref{tbl1w} the masses associated with
the different shells are listed.

At the end of the simulated evolution the model has accumulated
an explosion energy of approximately $0.6\times 10^{51}\,$erg.
The mass cut and thus initial baryonic mass of the neutron star
is 1.41$\,$M$_{\odot}$. The model fulfills fundamental constraints
for Type II SN nucleosynthesis \citep{hof96} because the ejected
mass having $\ye\lesssim 0.47$ is 
$\lesssim$10$^{-4}\,$M$_{\odot}$ (see Fig.~\ref{massye}) and thus
the overproduction of N=50 (closed neutron shell) nuclei
of previous explosion models does not occur. More than 83\% 
of the ejected mass in the convective bubble and early wind phase
(in total 0.03$\,$M$_{\odot}$ in this rather low-energetic 
explosion) have $\ye > 0.5$. 
The ejection of mostly p-rich matter is in agreement 
with 1D general relativistic SN
simulations with Boltzmann neutrino transport in which the 
explosion was launched by artificially enhancing the neutrino
energy deposition in the gain layer \citep{thi03,fro04}. 
The reason for the proton excess is the capture
of electron neutrinos and positrons on neutrons, which is 
favored relative to the inverse reactions because of the 
mass difference between neutrons and protons and because 
electron degeneracy becomes negligible in the neutrino-heated 
ejecta \citep{fro04,qia96}.

Although the explosion in the considered SN model
of \citet{jan03} was obtained by a regression from the most
accurate treatment of the neutrino transport, it not only
demonstrates the proximity of such accurate models to 
explosions, but also provides a consistent description 
of the onset of the SN explosion due to the convectively supported
neutrino-heating mechanism, and of the early SN evolution. The
properties of the resulting explosion are very interesting, including 
the conditions for nucleosynthesis.
The $\ye$ values of the ejecta should be rather
insensitive to the manipulation which enabled the explosion.
On the one hand the expansion velocities of the high-entropy
ejecta are still fairly low (less than a few 
$10^8\,$cm$\,$s$^{-1}$) when weak interactions freeze out, and
on the other hand the omitted velocity-dependent effects 
affect neutrinos and antineutrinos in the same way.

\subsection{Outflows in the Convective Bubble}

In order to calculate the nucleosynthesis it is necessary to have a
starting composition and the temperature-density ($T-\rho$) history of
the matter as it expands and is ejected from the supernova. Because
the matter is initially in nuclear statistical equilibrium, the
initial values of $\ye$, $T$, and $\rho$ determine the composition
which is just protons with a mass fraction $\ye$ and neutrons. We
are most interested in the innermost few hundredths to one tenth of a solar
mass to be ejected. This matter has an interesting history. It was
initially part of the silicon shell of the star, but fell in when the
core collapsed, passed through the SN shock and was photodisintegrated to
nucleons. Neutrino heating then raised the entropy and energy of the
matter causing it to convect. Eventually some portion of this matter
gained enough energy to expand and escape from the neutron star,
pushing ahead of it the rest of the star. As it cooled, the nucleons
reassembled first into helium and then into heavy elements.

\begin{figure}
\epsscale{1.5}
\plotone{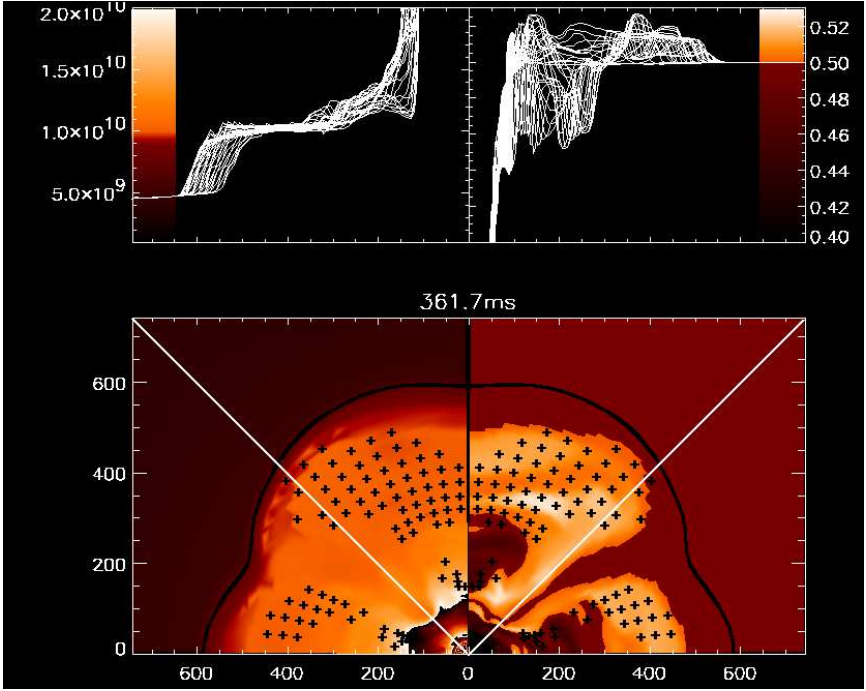}
\caption{Convective bubble around the central neutron star in 
the two-dimensional SN model of \citet{jan03}. Time in 
this figure is measured from the onset of collapse, core bounce
was at 175$\,$ms, the length scale is in km.
The panels on the left show the temperature
in Kelvin, those on the right $\ye$. The upper plots
give radial profiles for all angular bins of the polar
coordinate grid of the simulation, which was carried out in
a lateral wedge of $\pm 43.2^{\mathrm o}$ (with
periodic boundary conditions) around the equatorial plane. The
latter is indicated by white diagonal lines. The
positions of the tracer particles at the onset of the explosion
are marked by crosses in the lower panels. Their positions were
chosen (by post-processing the finished simulation) such that
the $\ye$ distribution of the final ejecta was appropriately
represented.
\label{fig0}}
\end{figure}

The temperature-density history of such matter is thus not given by
the simple ansatz often employed in explosive nucleosynthesis ---
``adiabatic expansion on a hydrodynamic time scale''. In fact, owing
to convection, the temperature history may not even be monotonic.
Here we rely on tracer particles embedded in the so called ``hot
convective bubble'' of the 15 $\Msun$ SN model calculated by
\citet{jan03} (Fig.~\ref{fig0}). These tracer particles were not
distributed uniformly in mass, but chosen to represent a range of
$\ye$ in the ejecta.

The proton-rich outflows of interest here begin at about 190 ms after
core bounce (Fig.~\ref{massshells}). 
Entropies and electron fractions characteristic of a few
different trajectories are given in Table \ref{tbl1}. Each trajectory
represents a different mass element in the convective bubble.  As is
seen, $\ye$ for the different trajectories lies in the range from
$0.5-0.546$, and the entropies per nucleon are modest, $s/k_b\sim 30-50$.
Figure~\ref{massye} shows the ejected mass versus $\ye$
during the convective phase of the SN explosion.

At the end of the 2D calculation of \citet{jan03}, the mass element in a
typical trajectory had reached a radius of about $2000$ \km (corresponding
to the time when the SN model was mapped from 2D to 1D and thus 
detailed information for the mass elements was lost).
Temperatures at this radius were typically $T_9\equiv T/10^9\,{\rm
K} \approx 4$--5, which is still hot enough that nuclei have not yet
completely re-assembled.  To follow the nucleosynthesis until all
nuclear reactions had frozen out it was necessary to extrapolate the
trajectories to low temperature.  In doing so, we assumed that the
electron fraction and entropy were constant during the extrapolated
portion of the trajectory. This should be valid because the number of
neutrino captures suffered by nuclei beyond $\sim 2000$ \km is small.

We considered two approximations to the expansion which should bracket
the actual behavior. The first assumes homologous expansion at a
velocity given by the Janka et al. calculation between 10 billion and
4 billion K. This ignores any deceleration experienced as the hot
bubble encounters the overlying star and is surely an underestimate of
the actual cooling time (though perhaps realistic for the
accretion-induced collapse of a bare white dwarf). In particular, we
estimated the homologous expansion time scale for each trajectory as
$\tau_{\rm hom}=
(t_{\mathrm{f}}-t_{\mathrm{i}})/\ln(\rho_{\mathrm{i}}/\rho_{\mathrm{f}})$ 
where the subscript $\mathrm{i}$
denotes the value of a quantity when $T_9=10$ and the subscript $\mathrm{f}$
denotes the value of a quantity at the last time given for the tracer
particle history ($t_{\mathrm{f}}\approx 436\,{\rm ms},\,
T_{9,{\mathrm{f}}}\approx
4$--5). Values of $\tau_{\rm hom}$ for different trajectories are given
in Table \ref{tbl1}.

The second approximation was an attempt to realistically represent
material catching up with the supernova shock.  This extrapolation is
based on smoothly merging the trajectories found in the calculations
of Janka et al. with those calculated for the inner zone of the same
15 $\Msun$ supernova by \citet{Woo95}. There are some differences. The
earlier study was in one dimension and the shock was launched
artificially using a piston. The kinetic energy at infinity of the
Woosley-Weaver model was $1.2 \times 10^{51}$ erg; that of the Janka et al.\
model was $0.6 \times 10^{51}$ erg. Still the calculations agreed roughly
in the temperature and density at the time when the evaluation of
tracer particles in the current 2D simulation was
stopped.  In order not to have discontinuities in the entropy at the
time when the two calculations are matched, the density in the previous
1D calculation is changed slightly. This merging of the late time
trajectories is expected to be reasonable because the shock evolution
at several seconds post core bounce is determined mostly by the
explosion energy.

We shall see in Sect.~\ref{sec:nucresults} that abundances of key
nuclei are particularly sensitive to the time it takes the flow to
cool from $2\cdot 10^9\,$K to $1\cdot 10^9\,$K. The homologous expansion
approximation gives this time as about 100--200 ms, while the Kepler
based estimate gives this time as about 1 sec. Both estimates are
rough and should be viewed as representing upper and lowed bounds to
the time scale.

\begin{figure}
\epsscale{1.5}
\plotone{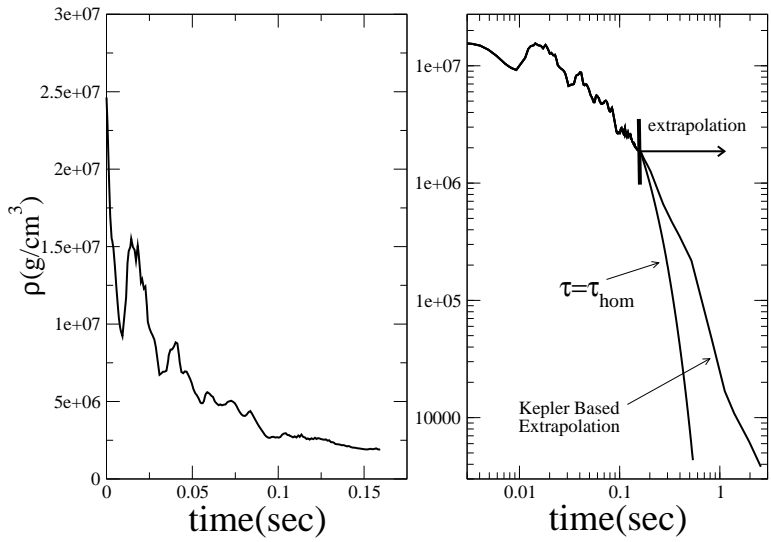}
\caption{Illustration of the density evolution for 
hot-bubble trajectory 10 
in the two-dimensional SN model of \citet{jan03}
(left panel) and as extrapolated for the present 
nucleosynthesis calculations (right panel). 
\label{figrho}}
\end{figure}

\begin{figure}
\epsscale{1.5}
\plotone{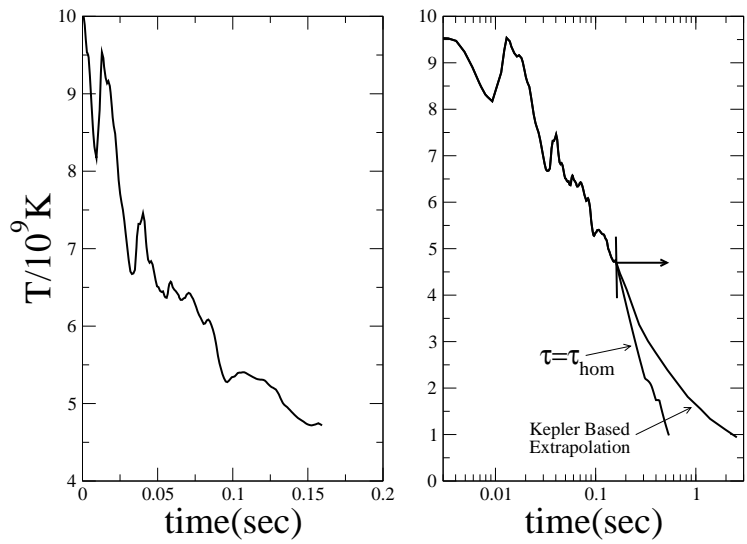}
\caption{Evolution of temperature in the trajectory displayed 
in Fig.~\ref{figrho}.
\label{figtemp}}
\end{figure}

Figure~\ref{figrho} shows the evolution of density in a representative
trajectory for each of the two approximations to the flow at large
radii. The temperature history in these trajectories is shown in
Fig.~\ref{figtemp}.  Note the irregular and non-monotonic evolution
of the thermodynamic quantities at early times.

\subsection{Outflows in the Early Wind}

While the shock sweeps through and expels the stellar mantle, matter
is still being continuously ablated from the surface of the cooling
neutron star. Neutrino heating, principally via charged current
neutrino capture, acts to maintain pressure-driven outflow in the
tenuous atmosphere formed by the ablated material. This outflow has a
higher entropy and is less irregular than the convective bubble.

The evolution of material at radii smaller than a few hundred km is
set by characteristics of the cooling neutron star. It is at these
small radii that the asymptotic entropy $s$ and electron fraction
$\ye$ are set. At early times the neutron star has yet to radiate away
the bulk of its gravitational energy and so has a relatively large
radius. Material escaping the star during this period only needs to
gain a little energy through heating to escape the still shallow
gravitational potential. Consequently, the entropy of the asymptotic
outflow is about a factor of two smaller than the entropy of winds
leaving the neutron star $\sim$10 seconds post core-bounce. This can
be seen from the analytic estimate provided by \cite{qia96}
\begin{equation}
s\approx 235 (L_{{\bar \nu}_e,51} \epsilon_{{\bar \nu}_e,{\rm
MeV}}^2)^{-1/6} \left(\frac{10^6\,\cm}{R}\right)^{2/3}.
\end{equation}
Here $L_{{\bar \nu}_e,51}=L_{\bar{\nu}_e}/10^{51}{\rm erg /sec}$ ,
$\epsilon_{{\bar \nu}_e,{\rm MeV}}$ approximately the mean energy of
electron anti-neutrinos and $R$ is the neutron star radius. A lower
entropy implies a higher density and therefore faster particle 
capture rates at a given
temperature. For proton-rich outflows this typically results in
synthesis of heavier elements.

The electron fraction in the outflow is set by a competition between 
different lepton capture processes on free nucleons:
\begin{eqnarray}
\nu_e+ {\rm n} & \longleftrightarrow & {\rm p} + e^{-} \, ,\\
e^+ + {\rm n} & \longleftrightarrow & {\rm p} + \bar{\nu}_e \, .
\end{eqnarray}
Because the neutron star is still deleptonizing at early times, the
$\nu_e$ and $\bar{\nu}_e$ spectra can be quite similar. Also, once
heating raises the entropy of material leaving the neutron star, the
number densities and spectra of electrons and positrons within the
material become similar. Under these circumstances the 1.29 MeV
threshold for ${\rm p\rightarrow n}$ results in $\bar{\nu}_e/e^-$ capture
rates which are slower than the inverse $\nu_e/e^+$ capture
rates. Weak processes then drive the outflow proton rich. 
The electron fraction in the wind is mostly set by the competition 
between $\nu_e$ and $\bar{\nu}_e$ capture (because $e^\pm$ captures
freeze out when the density and temperature in the outflow become low,
whereas high-energy neutrinos streaming out from the neutrinosphere 
still continue to react with nucleons). When the composition comes
to equilibrium with the neutrino fluxes,
\begin{equation}
\ye\,\approx\, \frac{\lambda_{\nu_e n}}{\lambda_{\nu_e n}
+ \lambda_{\bar{\nu}_e p}}\ .
\end{equation}
Here $\lambda_\nu$ represents the electron neutrino or antineutrino 
capture rate on neutrons or protons. Because the star is still 
deleptonizing at early times, the $\nu_e$ and $\bar{\nu}_e$ spectra
can be quite similar. The 1.29$\,$MeV threshold for $\bar{\nu}_e$
capture then leads to $\lambda_{\nu_e n} > \lambda_{\bar{\nu}_e p}$,
and proton-richness is established in the outflow. Finally
also the neutrino reactions cease because of
the $1/r^2$ dilution of the neutrino density with growing distance
$r$ from the neutron star.

Table \ref{tbl1w} gives characteristics of the early wind found in the
simulations of \citet{jan03}. As expected, the wind is proton rich at
early times. Eventually, the hardening of the
$\bar{\nu}_e$ spectrum relative to the $\nu_e$ spectrum will cause
$\ye$ to fall below 1/2. This turnover has not yet occurred when the
hydrodynamic simulation was stopped. It should take place at a later
time when the wind properties (mass loss rate, entropy) have changed
such that the nucleosynthesis constraints for the amount of 
$\ye< 0.47$ ejecta \citep{hof96} will not be violated. At 1.3$\,$s
after bounce the mass loss rate of about 
$3\times 10^{-3}\,$M$_{\odot}\,$s$^{-1}$ and wind entropy of
$\sim\,$80$\,k_b$ per nucleon in the Janka et al.\ model are likely
to still cause an overproduction of N=50 nuclei if $\ye$ went 
significantly below 0.5.

The temperature in the wind at the end of the traced shell expansion
is $T_9\approx 2$ (Fig.~\ref{wind}). Approximations 
for the wind evolution at lower temperatures are the same as those
discussed above.

\subsection{Nuclear Physics Employed}

The reaction network used for the present calculations is given in
Table \ref{reactable}.  Estimates of reaction rates and nuclear
properties used in our calculations are the same as those used in the
study of X-ray bursts by \cite{Woo04}. Briefly, reaction rates were
taken from experiment whenever possible, from detailed shell-model
based calculations \citep{Fis01} for a few key $({\rm p,\gamma})$
rates, and from Hauser-Feshbach calculations \citep{Rau00}
otherwise. Proton separation energies, which are crucial determinants
of nucleosynthesis in flows with ${\ye>1/2}$, were taken from a
combination of experiment \citep{Aud95}, the Hartree-Fock Coulomb
displacement calculations of \cite{bro02} for many important nuclei
with Z$>$N, and theoretical estimates \citep{Mol95}. Choosing the best
nuclear binding energies is somewhat involved and we refer the reader
to the discussion in \cite{bro02} and Fig.~1 of \cite{Woo04}.
Ground-state weak lifetimes are experimentally well
determined for the nuclei important in this paper. At temperatures
larger than $10^9$ K the influence of thermal effects on weak decays
was estimated from the compilation of \cite{Ful82} where
available. Table \ref{fultable} gives the nuclei for which the Fuller
et al. rates were used. A test calculation in which we switched
thermal rates off and used only experimentally determined ground-state
rates showed little effect on the important abundances. Section
\ref{sec:nuclear} contains a discussion of the influence of nuclear
uncertainties on yields of some interesting nuclei.

\section{Nucleosynthesis Results}
\begin{figure}
\epsscale{1.5}
\plotone{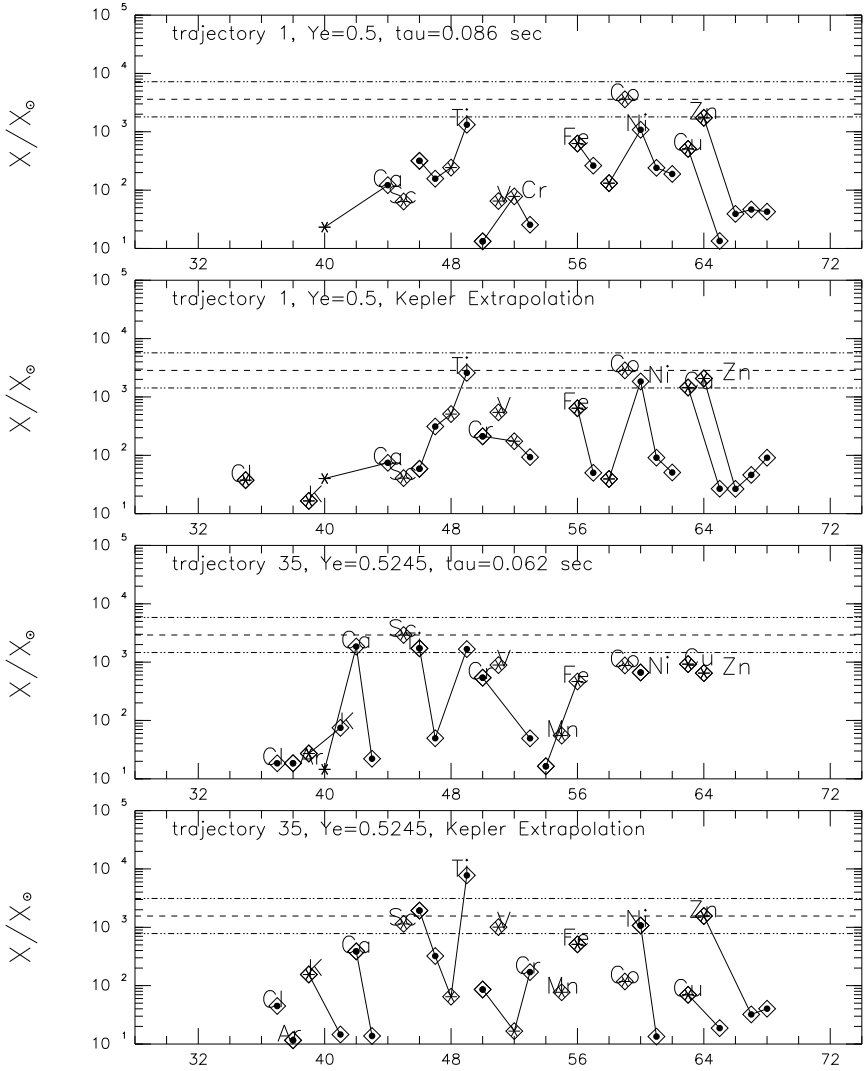}
\caption{Overproduction plot for nucleosynthesis in some of the 
tracer particle trajectories in the convective bubble of
the SN model of \citet{jan03}. Results for 
two assumptions about the dynamic time scale are shown for each trajectory.\label{fig2}}
\end{figure}

\lSect{nucresults} Table \ref{tbl1} gives the major calculated
production factors for a number of trajectories in the convective
bubble and for our two different estimates of the material expansion
rate at low temperatures. Table \ref{tbl1w} gives production factors
for nuclei synthesized in different mass elements comprising the early
wind.  Here the production factor for nuclide $i$ is defined as
\begin{equation}
P_i={M \over M^{\mathrm{ej}}}{X_i \over X_{\odot,i}},
\end{equation}
where $M$ is the total mass in a given trajectory, 
$M^{\mathrm{ej}}=13.5\,\Msun$
is the total mass ejected in the SN explosion, $X_i$ is 
the mass fraction of nuclide $i$ in the trajectory, and $X_{\odot,i}$
is the mass fraction of nuclide $i$ in the sun. 
To aid in interpreting the tables we show in Fig.~\ref{fig2}
plots of $X_i/X_{\odot,i}$ characterizing the
nucleosynthesis in two representative hot-bubble trajectories. 

Production factors integrated over the different bubble trajectories
are given in Table \ref{tbl2}. If one assumes rapid expansion,
production factors of $^{45}$Sc, $^{63}$Cu, $^{49}$Ti, and $^{59}$Co
are all above 1.5.  For the slower expansion time scale below $4
\times 10^9$ K, which we regard as more realistic, a different set of
nuclei are produced, especially $^{49}$Ti and $^{64}$Zn. Depending
upon mass and metallicity, $^{49}$Ti may already be well produced in
other regions of the same supernova \citep{Woo95,Rau02}, but $^{64}$Zn
is not.  The synthesis here thus represents a new way of making
$^{64}$Zn and this same process will function as well in zero and low
metallicity stars as in supernovae today. However, $^{64}$Zn was
already known to be produced, probably in greater quantities, by the
neutrino-powered wind \citep{hof96}.

Production factors integrated over the different wind trajectories are
given in Table \ref{tbl2w}. The somewhat high-entropy wind synthesizes
$^{45}$Sc, $^{49}$Ti and $^{46}$Ti more efficiently than the
bubble. Typical values of $X/X_{\sun}$ for these three nuclei are
approximately $10^4$ in the wind and approximately $2\cdot 10^3$ in
the bubble. In the present calculations the integrated production
factor for Sc in the wind is between about 1.5 and 4.7 depending on
the time scale describing the wind expansion at $T_9\lesssim 2$.

For comparison, in the 15 $\Msun$ supernova of \citet{Rau02}, this
production factor was about 7 for many major species, including
oxygen. This is close to the combined wind/bubble production factors
of Sc and $^{46,49}$Ti in the present calculations. The other most
abundant productions in Tables \ref{tbl2} and \ref{tbl2w} fall short
of this - but not by much. The bulk production factors in a 25 $\Msun$
supernova are about twice those in a 15, but our explosion model is
not easily extrapolable to stars of other masses. {\sl If} 25 $\Msun$
stars explode with a similar kinetic energy it will probably take a
more powerful central engine to overcome their greater binding energy
and accretion rate during the explosion.  Probably this requires more
mass in the convective bubble. In fact, the energy of the 15 $\Msun$
supernova used here, $0.6 \times 10^{51}$ erg, would be regarded by
many as low. It may be that the mass here should be doubled too.

It is important to note that the species listed in Tables \ref{tbl2}
and \ref{tbl2w} are not made as themselves but as proton-rich
radioactive progenitors.  Major progenitors of important product
nuclei are given in the far right column of Table \ref{tbl2}.
Typical progenitors of important nuclei are 3--4 charge units from
stability.  This can be understood through consideration of the Saha
equation. Before charged particle reactions freeze out at $T_9\approx
1.5-2$, nuclear abundances along an isotonic chain are well
approximated as being in local statistical equilibrium:
\begin{equation}
\label{localsaha}
\frac{X({\rm Z+1,N})}{X({\rm Z,N})} \approx 10^{-5} \exp(S_p/T) \frac{\rho_5}{T_9^{3/2}}\frac{G_{\rm Z+1,N}}{G_{\rm Z,N}}.
\end{equation}
Here $S_p$ is the proton separation energy of the Z+1,N nuclide, $G$
represents the partition function, $\rho_5=\rho/10^5{\gcc}$,
$T_9=T/10^9{\rm K}$, and A=Z+N. Equation~(\ref{localsaha}) predicts 
that the abundances of nuclei with $S_p\lesssim 500$ keV are very small.

Perhaps the most notable feature of the proton-rich trajectories is
their inefficiency at synthesizing elements with
A$\,\gtrsim\,$60. Neutron-rich outflows, by contrast, readily synthesize
nuclides with mass A$\,\sim\,$100. This is shown in Table \ref{tbl3}
which gives production factors characterizing nucleosynthesis in
somewhat neutron-rich winds occurring in the SN. The Kepler-based
extrapolation of the first trajectory in Table \ref{tbl1} is used for
these $\ye<0.5$ calculations. Estimates of the mass in each $\ye$ bin
for the calculations of \cite{jan03} are shown in Fig.~\ref{massye}.

Termination of the nuclear flow at low mass number in proton-rich outflows
has a simple explanation. Unlike nuclei at the neutron drip lines,
proton-rich waiting point nuclei have lifetimes much longer than the
time scales characterizing expansion of neutrino-driven outflows. In
addition, proton capture from waiting point nuclei to more rapidly
decaying nuclei is inefficient. To illustrate the difficulty with
rapidly assembling heavier proton-rich nuclei, consider nuclear flow
through $^{64}{\rm Ge}$.  This waiting point nucleus has a lifetime of
approximately 64 sec.  The ratio of the amount of flow leaving
$^{65}{\rm As}$ to that leaving $^{64}{\rm Ge}$ is found from
application of the Saha equation above,
\begin{equation}
\frac{\lambda_+(^{65}{\rm As})Y(^{65}{\rm As})}{\lambda_+(^{64}{\rm
Ge})Y(^{64}{\rm Ge})} \approx 10^{-2} \frac{\rho_5}{T_9^{3/2}} \exp(S_p/T).
\end{equation}
Here $\lambda_+$ represents the $\beta^+$ decay rate and $S_p$ is the
proton separation energy of As. For $^{65}{\rm As}$, $\lambda_+\approx
\ln(2)/0.1\sec$ and for $^{64}{\rm Ge}$, $\lambda_+\approx
\ln(2)/64\sec$. By definition, proton capture daughters of waiting
point nuclei are characterized by small proton separation
energies. The binding energy of $^{65}{\rm As}$ still has large
uncertainties, though is known to be less than about 200keV
\citep{bro02}. Positron decay out of the proton capture daughter of
the waiting point nuclei is negligible for such small proton
separation energies.  These considerations do not hold for X-ray
bursts, where time scales characterizing nuclear burning can be tens or
hundreds of seconds.

\begin{figure}
\epsscale{1.5}
\plotone{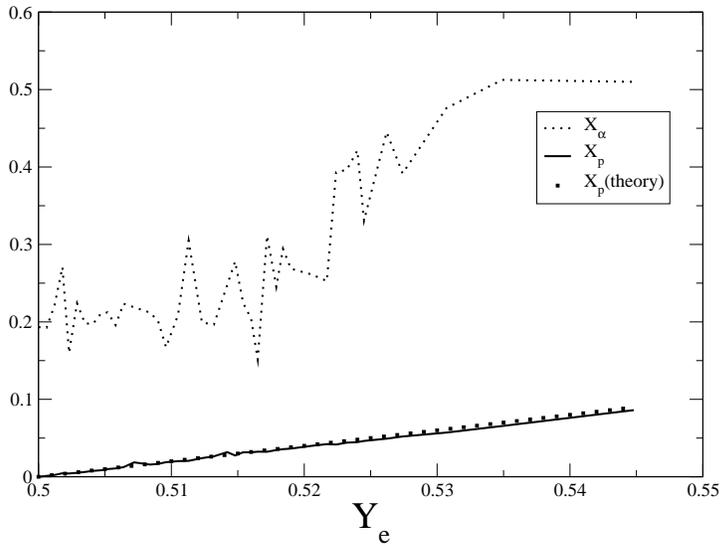}
\caption{Final proton and $\alpha$ particle mass fractions as a function 
of $\ye$ for the different Kepler extrapolated trajectories. 
The line labeled with $X_p({\rm theory})$ is the 
proton mass fraction calculated under the assumption that all available nucleons are locked into alpha 
particles or other nuclei with equal and even numbers of protons and neutrons.\label{xpandxalpha}
}
\end{figure}

The difficulty with rapid assembly of heavy proton-rich nuclei is also
evident in the final free proton and alpha particle mass fractions.
The trend of $X_p$ and $X_{\alpha}$ with $\ye$ is shown in 
Fig.~\ref{xpandxalpha} for the different Kepler extrapolated bubble
trajectories.  Also shown in this figure is the proton mass fraction
calculated under the assumption that all available nucleons are bound
into alpha particles. This is an approximate measure of the mass
fraction of available protons. Note that the mass fraction of protons
in the two calculations are nearly identical. This is because assembly
of proton-rich nuclei occurs on a very slow time scale set by a few
$\beta^+$ rates.

Because nucleosynthesis past A$\,\sim\,$60 is inefficient these
proton-rich flows do not produce N=50 closed shell
nuclei. Historically, overproduction of N=50 nuclei has plagued
calculations of supernova nucleosynthesis
\citep{How93,Wit93,woo94}. The influence of weak interactions in
driving some of the outflow to ${\ye >1/2}$ ameliorates this problem.

\subsection{Details of the Nucleosynthesis and Critical Nuclear Physics}
\lSect{nuclear}

\begin{figure}
\epsscale{1.5}
\plotone{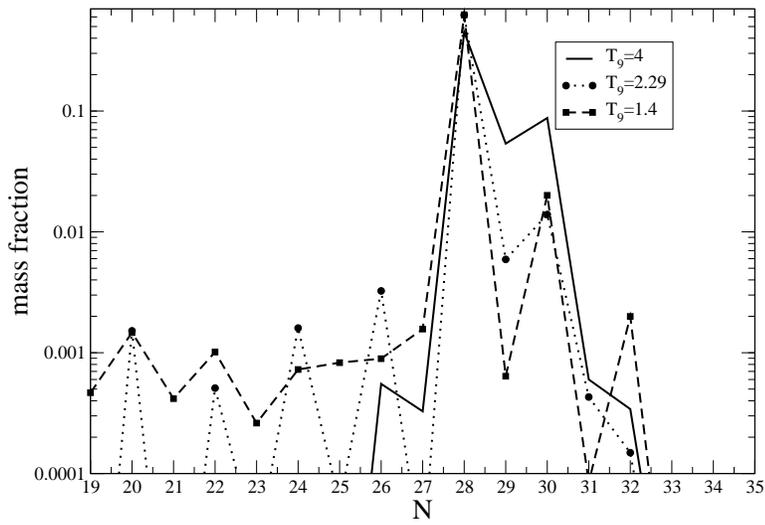}
\caption{Evolution of nuclear abundances as a function of neutron 
number in the Kepler based extrapolation for the hot-bubble
trajectory 26 ($\ye=0.5172$).\label{nfig}}
\end{figure}

To aid in understanding the general character of these proton-rich
flows we show in Fig.~\ref{nfig} the evolution of nuclear mass
fractions as a function of the neutron number. At $T_9\approx 4$,
$\alpha$ captures have led to efficient synthesis of tightly bound
species with N=28 and N=30. As temperature decreases $\alpha$ capture
becomes less efficient and $\beta^+$ decay drives flow to higher
neutron number. From Table~\ref{tbl2} it is seen that the nuclei we
are most interested in arise from decay of nuclei with N=21, 24, 31
and 32. From Fig.~\ref{nfig} it is clear that synthesis of nuclei
with these neutron numbers represents a minor perturbation on the
nucleosynthesis as a whole.

Tables \ref{tbl2} and \ref{tbl2w} show that $^{45}{\rm Sc}$, the only
stable scandium isotope, has a combined wind/bubble 
production factor of about 6 if
freeze-out is rapid and a combined production factor about 50$\%$ smaller in
the slower Kepler extrapolated trajectories. 
Efficient synthesis of
scandium in proton rich outflows associated with Gamma Ray Bursts has
been noted previously by \cite{pru04}, while \cite{mae03} found that
scandium may also be synthesized explosively in shocks exploding
anomalously energetic supernovae. Indeed, values presented here for
${\ye}$, $s/k_b$, and $\tau$ in the early SN wind are very close to
estimates of these quantities in winds leaving the inner regions of
accretion disks powering collapsars \citep{mac99,pru204}.  The origin
of Sc is currently uncertain and it may be quite abundant in low
metallicity stars \citep{Cay04} suggesting a primary origin.  In the
present calculations the yields of this element are close to those
needed to explain the current inventory of Sc.

To understand how synthesis of scandium depends on the outflow
parameters and nuclear physics, note that Sc arises mostly from
$\beta^+$ decay originating with the quasi waiting-point nucleus
$^{45}{\rm Cr}$. In turn, N=21 isotones of $^{45}{\rm Cr}$ originate
from $\beta^+$ decay out of isotones of $^{40}{\rm Ca}$. The doubly
magic nucleus $^{40}{\rm Ca}$ is efficiently synthesized through a
sequence of alpha captures.  At temperatures larger than about $2\cdot
10^9$ K statistical equilibrium keeps almost all N=20 nuclei locked
into $^{40}{\rm Ca}$. This nucleus is $\beta$ stable and has a first
excited state at 3.3 MeV, too high to be thermally populated. Flow out
of N=20 can only proceed when the temperature drops to approximately
1.5 billion degrees and statistical equilibrium favors population of
$^{42}{\rm Ti}$ over $^{40}{\rm Ca}$. The proton capture daughter of
$^{40}{\rm Ca}$ ($^{41}{\rm Sc}$) has a proton separation energy of
only 1.7 MeV and is not appreciably abundant. Decay out of $^{42}{\rm
Ti}$ is then responsible for allowing flow to N=21. $^{42}{\rm Ti}$
has a well determined $\beta^+$ half life of 199$\pm$6 ms, a proton
separation energy which is uncertain only by about 5 keV, and a first
excited state too high in excitation energy to play a role in allowing
flow to N=21. In short, nuclear properties are well determined for
important N=20 nuclei. Once nuclei make their way to N=21 at
$T_9\approx 1.5$, their abundances are divided between the tightly
bound $^{45}{\rm Cr}$ and $^{43}{\rm Ti}$. Here uncertainties in
nuclear physics may be more important. For $^{45}{\rm Cr}$ the proton
separation energy is uncertain to about 100 keV and the spin of the
ground state is uncertain. To the extent that the relative abundances
are set by the Saha equation, these uncertainties could imply an
uncertainty of a factor of several in the relative abundances of
$^{45}{\rm Cr}$ and $^{45}{\rm Ti}$ at $T_9\approx 1.5$. In turn, this
implies appreciable uncertainty in the estimated Sc yield.

Whether or not Sc is efficiently synthesized following decay of
$^{45}$Cr depends on the expansion time scale at low temperatures. This
is because the $\beta^+$ daughter of $^{45}$Cr is $^{45}$V, which has
a relatively small proton separation energy of 1.6 MeV. At low
temperatures the Saha equation favors proton capture to $^{46}$Cr. If
the expansion is slow enough that most $^{45}$Cr decays at
temperatures where $^{45}{\rm V(p,\gamma)^{46}Cr}$ is still rapid,
then flow out of the N=22 nuclei occurs via $\beta^+$ decay out of
$^{46}$Cr. In this case $^{46}$Ti is synthesized rather than
$^{45}$Sc.

$^{49}{\rm Ti}$ originates from the the N=24 nuclide $^{49}{\rm Mn}$.
At $T_9\approx 1.4$ nuclei with N=24 are divided roughly equally
between $^{49}{\rm Mn}$ and $^{50}{\rm Fe}$.  Uncertainties in the
proton separation energies and lifetimes of these nuclei are small.
$^{49}{\rm Mn}$ does have a low lying excited state at 382$\,$keV
which is thermally populated at low temperatures. However, $^{49}{\rm
Mn}$ is a nucleus with Z=N+1 that is expected to have ground and excited
state decay rates that are dominated by super-allowed Fermi
transitions which are almost independent of excitation energy.

Lastly, we turn our attention to flow out of the N=32 isotones which
are progenitors of $^{60}{\rm Zn}$ and $^{63}{\rm Cu}$. Proton-rich
nucleosynthesis near $^{64}{\rm Ge}$ has been extensively discussed in
the X-Ray Burst literature (e.g. Brown et al.\ 2002). 
Uncertainties in basic
nuclear properties important for synthesis of $^{60}{\rm Zn}$ are
small. This is not true for $^{63}{\rm Cu}$, which is formed directly
by the decay of $^{63}{\rm Ga}$. $^{63}{\rm Ga}$ has a $J^\pi=(5/2)^-
$excited state at 75.4 keV which dominates the partition function at
$T_9\approx 1.5$ since the ground state has $J=3/2$. The weak lifetime
of this excited state is experimentally undetermined (as are the weak
lifetimes of all short lived excited states) and could easily be a
factor of five longer or shorter than the quite long ground state
lifetime of $\sim 32$ sec. This translates into an uncertainty of a
factor of several in the inferred $^{63}{\rm Cu}$ yield.

The influence of possible uncertainties in the time scale, entropy, and
electron fraction characterizing the different trajectories can be
seen from the results in Table \ref{tbl1}. Modest changes in the
outflow parameters result in factors of $\sim 2$ changes in yields of
the most important isotopes. This is evident by the quite different
efficiencies with which the lower entropy bubble and higher entropy
wind synthesize $^{45}$Sc and $^{49}$Ti.

So far we have not considered the influence of neutrino interactions,
except implicitly through the setting of $\ye$. If matter remains
close to the neutron star, neutrino capture and neutrino-induced
spallation may compete with positron decay, even on a dynamic
time scale. However, neutrino capture alone cannot act to accelerate
nuclear flow past waiting point nuclei and allow synthesis of the
heavier proton-rich elements. The reason is that the neutrino capture
rates on the waiting point nuclei are about the same as the rate of
neutrino capture on a free proton \citep{woo90}. Every capture of a
neutrino by a heavy nucleus is accompanied by a capture onto a free
proton. The electron fraction is then rapidly driven to $1/2$ since
the neutron produced in this way immediately goes into the formation
of an $\alpha$-particle. This is analogous to the ``$\alpha$-effect''
discussed in the context of late-time winds \citep{ful95,mey98}.

\section{Conclusions and Implications}

The important news is that, unlike simulations of a few years ago,
there is no poisonous overproduction of neutron-rich nuclei in the
vicinity of the N = 50 closed shell \citep{woo94}. When followed in
more detail (i.e.\ mainly with a better, spectral treatment of the
neutrino transport), weak interactions in the hot convective bubble drive
$\ye$ back to 0.5 and above so that most of the mass comes out as
$^{56}$Ni and $^4$He. Since $^{56}$Fe and helium are abundant in
nature, this poses no problem.

Beyond this it is also interesting that the proton-rich environment of
the hot convective bubble and early neutrino-driven wind
can synthesize interesting amounts of some
comparatively rare intermediate mass elements. If the total mass of
SN ejecta with $\ye\gtrsim 0.5$ is larger than a few hundredths of a
solar mass, these proton-rich outflows may be responsible for a
significant fraction of the solar abundances of $^{45}{\rm Sc}$,
$^{64}{\rm Zn}$, and some Ti isotopes, especially $^{49}$Ti.

However, these ejecta do not appear to be implicated in the synthesis
of elements that do not have other known astrophysical production
sites. For example, ${\rm Sc}$ can be produced explosively, while
$^{64}{\rm Zn}$ can be synthesized in a slightly neutron-rich wind.
It seems unlikely that consideration of nucleosynthesis in proton-rich
outflows will lead to meaningful constraints on conditions during the
early SN.

Since the conditions in the hot convective bubble resemble in some
ways those of Type I X-ray bursts (high temperature and proton mass
fraction), we initially hoped that the nuclear flows would go higher,
perhaps producing the $p$-process isotopes of Mo and Ru. Such species
have proven difficult to produce elsewhere and the $rp$-process in
X-ray bursts can go up as high as tellurium
\citep{Sch01}. Unfortunately the density is much less here than in the
neutron star and the time scale shorter. Proton-induced flows are
weaker and the leakage through critical waiting point nuclei is
smaller. Using the present nuclear physics, significant production
above A$\,$=$\,$64 is unlikely. However, heavier nuclei can be produced in
ejecta that are right next to these zones but with values of $\ye$
considerably less than 0.50 \citep{hof96}.

\acknowledgments 

This work was performed under the auspices of the
U.S. Department of Energy by University of California Lawrence
Livermore Laboratory under contract W-7405-ENG-48.
HTJ enjoyed discussions with Matthias Liebend\"orfer.
RB and HTJ thank A.~Marek for assistance with data 
evaluation and visualization, and
acknowledge support by the Son\-der\-for\-schungs\-be\-reich
375 ``Astro-Particle Physics'' of the Deut\-sche 
For\-schungs\-ge\-mein\-schaft. The supernova simulations were
done at the Rechenzentrum Garching and at the John von Neumann
Institute for Computing (NIC) in J\"ulich.

\clearpage
\begin{deluxetable}{cccccccccc}
\tabletypesize{\scriptsize}
\tablecaption{Reaction Network Used for the Present Calculations
\label{reactable}}
\tablewidth{0pt}
\tablehead{
\colhead{Element} &
\colhead{${\rm N_{\rm min}}$\tablenotemark{a}} &
\colhead{${\rm N_{\rm max}}$\tablenotemark{b}} &
\colhead{Element} &
\colhead{${\rm N_{\rm min}}$\tablenotemark{a}} &
\colhead{${\rm N_{\rm max}}$\tablenotemark{b}}&
\colhead{Element} &
\colhead{${\rm N_{\rm min}}$\tablenotemark{a}} &
\colhead{${\rm N_{\rm max}}$\tablenotemark{b}}&
}
\startdata
H  &  1  &  2  & He  &  1  &  4 & Li  &  3  &  6  \\
Be  &  3  &  8  & B  &  3  &  9 & C  &  3  &  12  \\
N  &  4  &  14  & O  &  5  &  14 & F  &  5  &  17  \\
Ne  &  6  &  21 & Na  &  6  &  33 & Mg  &  6  &  35  \\
Al  &  7  &  38 & Si  &  8  &  40 & P  &  8  &  42  \\
S  &  8  &  44 & Cl  &  8  &  46 & Ar  &  9  &  49  \\
K  &  11  &  51 & Ca  &  10  &  53 & Sc  &  13  &  55  \\
Ti  &  12  &  58 & V  &  15  &  60 & Cr  &  14  &  62  \\
Mn  &  17  &  64 & Fe  &  16  &  66 & Co  &  19  &  69  \\
Ni  &  18  &  71 & Cu  &  21  &  73 & Zn  &  21  &  75  \\
Ga  &  24  &  77 & Ge  &  23  &  80 & As  &  26  &  82  \\
Se  &  25  &  84 & Br  &  28  &  86 & Kr  &  27  &  88  \\
Rb  &  31  &  91 & Sr  &  30  &  93 & Y  &  33  &  95  \\
Zr  &  32  &  97 & Nb  &  35  &  99 & Mo  &  35  &  102  \\
Tc  &  38  &  104 &  Ru  &  37  &  106 & Rh  &  40  &  108  \\
Pd  &  40  &  110 & Ag  &  41  &  113 & Cd  &  42  &  115  \\
In  &  43  &  117 & Sn  &  44  &  119 & Sb  &  46  &  120  \\
Te  &  47  &  121  
\enddata
\tablenotetext{a}{Minimum neutron number included for the given element.}
\tablenotetext{b}{Maximum neutron number included for the given element.}
\end{deluxetable}

\clearpage
\begin{deluxetable}{ccccccc}
\tabletypesize{\scriptsize}
\tablecaption{Characteristics of nucleosynthesis in some representative bubble trajectories
\label{tbl1}}
\tablewidth{0pt}
\tablehead{
\colhead{Trajectory} &
\colhead{$\ye$} &
\colhead{$s/k_b$} &
\colhead{${\tau_{\rm hom}}$ (sec)} &
\colhead{$m/\Msun$} &
\colhead{Production ($\tau_{\rm dyn}={\tau_{\rm hom}}$)\tablenotemark{a}} &
\colhead{Production (Kepler Based Extrapolation)\tablenotemark{a}}}
\startdata
 1 & 0.500 & 18.4  & 0.086 & 9.25e-04  & $^{59}{\rm Co}$(0.33) & $^{64}{\rm Zn}$(0.17)\\
    &       &       &       &           & $^{64}{\rm Zn}$(0.12) & $^{59}{\rm Co}$(0.17)\\
    &       &       &       &           & $^{49}{\rm Ti}$(0.10) & $^{49}{\rm Ti}$(0.16)\\
 5 & 0.502 & 15.9  & 0.066 & 7.05e-04  & $^{59}{\rm Co}$(0.17) & $^{64}{\rm Zn}$(0.30)\\
    &       &       &       &           & $^{63}{\rm Cu}$(0.15) & $^{49}{\rm Ti}$(0.14)\\
    &       &       &       &           & $^{49}{\rm Ti}$(0.12) & $^{60}{\rm Ni}$(0.09)\\
 10 & 0.505 & 21.7  & 0.062 & 3.58e-04  & $^{59}{\rm Co}$(0.07) & $^{64}{\rm Zn}$(0.10)\\
    &       &       &       &           & $^{63}{\rm Cu}$(0.05) & $^{49}{\rm Ti}$(0.09)\\
    &       &       &       &           & $^{49}{\rm Ti}$(0.05) & $^{46}{\rm Ti}$(0.04)\\
 20 & 0.513 & 17.8  & 0.104 & 4.63e-04  & $^{45}{\rm Sc}$(0.14) & $^{49}{\rm Ti}$(0.19)\\
    &       &       &       &           & $^{46}{\rm Ti}$(0.06) & $^{64}{\rm Zn}$(0.07)\\
    &       &       &       &           & $^{42}{\rm Ca}$(0.05) & $^{60}{\rm Ni}$(0.05)\\
 30 & 0.521 & 26.2  & 0.047 & 2.67e-04  & $^{59}{\rm Co}$(0.03) & $^{49}{\rm Ti}$(0.08)\\
    &       &       &       &           & $^{45}{\rm Sc}$(0.02) & $^{64}{\rm Zn}$(0.03)\\
    &       &       &       &           & $^{63}{\rm Cu}$(0.02) & $^{60}{\rm Ni}$(0.02)\\
 35 & 0.524 & 26.9  & 0.062 & 2.28e-04  & $^{45}{\rm Sc}$(0.07) & $^{49}{\rm Ti}$(0.13)\\
    &       &       &       &           & $^{42}{\rm Ca}$(0.04) & $^{46}{\rm Ti}$(0.03)\\
    &       &       &       &           & $^{46}{\rm Ti}$(0.03) & $^{64}{\rm Zn}$(0.03)\\
  40 & 0.545 & 40.6  & 0.024 & 3.12e-04  & $^{42}{\rm Ca}$(0.04) & $^{49}{\rm Ti}$(0.25)\\
    &       &       &       &           & $^{45}{\rm Sc}$(0.04) & $^{46}{\rm Ti}$(0.06)\\
    &       &       &       &           & $^{46}{\rm Ti}$(0.03) & $^{45}{\rm Sc}$(0.04)
\enddata
\tablenotetext{a}{$\,\!$Listed here are the three nuclei with the largest
production factors. The production factor
for each nucleus is given in parenthesis next to the nucleus.}
\end{deluxetable}

\clearpage
\begin{deluxetable}{ccccccc}
\tabletypesize{\scriptsize}
\tablecaption{Production factors in the early wind
\label{tbl1w}}
\tablewidth{0pt}
\tablehead{
\colhead{$\ye$} &
\colhead{$s/k_b$} &
\colhead{${\tau_{\rm hom}}$ (sec)} &
\colhead{$m/\Msun$} &
\colhead{Production ($\tau_{\rm dyn}={\tau_{\rm hom}}$)\tablenotemark{a}} &
\colhead{Production (Kepler Based Extrapolation)\tablenotemark{a}}}
\startdata
0.551 & 54.8  & 0.131 & 1.53e-03  & $^{45}{\rm Sc}$(1.73) & $^{49}{\rm Ti}$(2.02)\\
         &       &       &           & $^{49}{\rm Ti}$(0.97) & $^{46}{\rm Ti}$(0.70)\\
           &       &       &           & $^{46}{\rm Ti}$(0.87) & $^{45}{\rm Sc}$(0.36)\\
 0.558 & 58.0  & 0.127 & 6.40e-04  & $^{45}{\rm Sc}$(0.95) & $^{49}{\rm Ti}$(1.09)\\
           &       &       &           & $^{49}{\rm Ti}$(0.52) & $^{46}{\rm Ti}$(0.38)\\
           &       &       &           & $^{46}{\rm Ti}$(0.48) & $^{45}{\rm Sc}$(0.20)\\
 0.559 & 76.7  & 0.099 & 6.80e-04  & $^{45}{\rm Sc}$(0.60) & $^{49}{\rm Ti}$(1.07)\\
          &       &       &           & $^{49}{\rm Ti}$(0.38) & $^{46}{\rm Ti}$(0.41)\\
           &       &       &           & $^{46}{\rm Ti}$(0.31) & $^{45}{\rm Sc}$(0.22)\\
 0.560 & 71.0  & 0.112 & 4.80e-04  & $^{45}{\rm Sc}$(0.55) & $^{49}{\rm Ti}$(0.79)\\
         &       &       &           & $^{49}{\rm Ti}$(0.31) & $^{46}{\rm Ti}$(0.29)\\
          &       &       &           & $^{46}{\rm Ti}$(0.27) & $^{45}{\rm Sc}$(0.15)\\
  0.568 & 74.9  & 0.059 & 8.00e-04  & $^{45}{\rm Sc}$(0.55) & $^{49}{\rm Ti}$(1.25)\\
          &       &       &           & $^{46}{\rm Ti}$(0.35) & $^{46}{\rm Ti}$(0.47)\\
          &       &       &           & $^{49}{\rm Ti}$(0.35) & $^{45}{\rm Sc}$(0.25)\\
 0.570 & 76.9  & 0.034 & 1.04e-03  & $^{46}{\rm Ti}$(0.38) & $^{49}{\rm Ti}$(1.49)\\
           &       &       &           & $^{45}{\rm Sc}$(0.35) & $^{46}{\rm Ti}$(0.57)\\
           &       &       &           & $^{42}{\rm Ca}$(0.31) & $^{45}{\rm Sc}$(0.31)
\enddata
\tablenotetext{a}{$\,\!$Listed here are the three nuclei with the largest
production factors. The production factor
for each nucleus is given in parenthesis next to the nucleus.}
\end{deluxetable}

\begin{deluxetable}{cccc}
\tabletypesize{\scriptsize}
\tablecaption{Production factors integrated over the different bubble trajectories
\label{tbl2}}
\tablewidth{0pt}
\tablehead{
\colhead{nucleus} &
\colhead{Production ($\tau_{\rm dyn}={\tau_{\rm hom}}$)} &
\colhead{Production (Kepler Based Extrapolation)} &
\colhead{Major Progenitor(s)}}
\startdata

  $^{59}{\rm Co}$ & 2.81 & 0.37 & $^{59}{\rm Cu},^{59}{\rm Zn}$ \\
  $^{49}{\rm Ti}$ & 2.00 & 6.53 & $^{49}{\rm Mn}$ \\
  $^{63}{\rm Cu}$ & 1.91 & 0.28 & $^{63}{\rm Ga},^{63}{\rm Ge}$ \\
  $^{45}{\rm Sc}$ & 1.65 & 1.33 & $^{45}{\rm Cr}$ \\
  $^{64}{\rm Zn}$ & 1.28 & 3.61 & $^{64}{\rm Ge}$ \\
  $^{46}{\rm Ti}$ & 1.22 & 1.97 & $^{46}{\rm Cr}$ \\
  $^{60}{\rm Ni}$ & 1.10 & 1.81 & $^{60}{\rm Zn}$ \\
  $^{42}{\rm Ca}$ & 1.04 & 0.46 & $^{42}{\rm Ti}$ 
\enddata
\end{deluxetable}

\begin{deluxetable}{ccc}
\tabletypesize{\scriptsize}
\tablecaption{Integrated production factors for the early wind
\label{tbl2w}}
\tablewidth{0pt}
\tablehead{
\colhead{nucleus} &
\colhead{Production ($\tau_{\rm dyn}={\tau_{\rm hom}}$)} &
\colhead{Production (Kepler Based Extrapolation)}}
\startdata
  $^{45}{\rm Sc}$ & 4.74 & 1.50\\
  $^{49}{\rm Ti}$ & 2.83 & 7.70\\
  $^{46}{\rm Ti}$ & 2.66 & 2.81\\
  $^{42}{\rm Ca}$ & 2.16 & 0.46\\
  $^{51}{\rm V\ }$ & 1.09 & 0.90\\
  $^{50}{\rm Cr}$ & 0.56 & 0.09
\enddata
\end{deluxetable}

\begin{deluxetable}{cccc}
\tabletypesize{\scriptsize}
\tablecaption{Characteristics of Nucleosynthesis in Neutron Rich Trajectories
\label{tbl3}}
\tablewidth{0pt}
\tablehead{
\colhead{$\ye$} &
\colhead{$m/\Msun$} &
\colhead{Production\tablenotemark{a}}}
\startdata
 0.470 & 6.40e-05 & $^{74}{\rm Se}$(6.59)\\
       &             & $^{78}{\rm Kr}$(4.25)\\
       &             & $^{64}{\rm Zn}$(1.35)\\
 0.475 & 7.98e-05 & $^{64}{\rm Zn}$(1.36)\\
       &             & $^{74}{\rm Se}$(0.85)\\
       &             & $^{78}{\rm Kr}$(0.78)\\
 0.480 & 1.59e-04 & $^{64}{\rm Zn}$(1.49)\\
       &             & $^{78}{\rm Kr}$(0.34)\\
       &             & $^{68}{\rm Zn}$(0.30)\\
 0.485 & 3.36e-04 & $^{62}{\rm Ni}$(0.92)\\
       &             & $^{58}{\rm Ni}$(0.35)\\
       &             & $^{64}{\rm Zn}$(0.23)\\
 0.490 & 6.24e-04 & $^{62}{\rm Ni}$(1.21)\\
       &             & $^{58}{\rm Ni}$(0.42)\\
       &             & $^{66}{\rm Zn}$(0.13)\\
 0.495 & 1.36e-03 & $^{62}{\rm Ni}$(1.30)\\
       &             & $^{58}{\rm Ni}$(0.41)\\
       &             & $^{61}{\rm Ni}$(0.23)
\enddata
\tablenotetext{a}{$\,\!$Listed here are the three nuclei with the largest
production factors. The production factor
for each nucleus is given in parenthesis next to the nucleus.}
\end{deluxetable}

\begin{deluxetable}{ccc}
\tabletypesize{\scriptsize}
\tablecaption{Nuclei for which thermal weak rates are included
\label{fultable}}
\tablewidth{0pt}
\tablehead{
\colhead{Atomic mass} &
\colhead{Elements\tablenotemark{a}}}
\startdata
 21 & F,  Mg,  Na,  Ne,  O \\
 22 & Mg,  Na,  Ne \\
 23 & F,  Mg,  Na,  Ne \\
 24 & Mg,  Na,  Ne,  Si \\
 25 & Mg,  Na,  Ne,  Si \\
 26 & Mg,  Na,  Si \\
 27 & Mg,  Na,  P,  Si \\
 28 & Mg,  Na,  P,  S,  Si \\
 29 & Mg,  Na,  P,  S,  Si \\
 30 & P,  S,  Si \\
 31 & Cl,  P,  S,  Si \\
 32 & Cl,  P,  S,  Si \\
 33 & Cl,  P,  S,  Si \\
 34 & Cl,  P,  S,  Si \\
 35 & Cl,  K,  P,  S \\
 36 & Ca,  Cl,  K,  S \\
 37 & Ca,  Cl,  K,  S \\
 38 & Ca,  Cl,  K,  S \\
 39 & Ca,  Cl,  K \\
 40 & Ca,  Cl,  K,  Sc,  Ti \\
 41 & Ca,  Cl,  K,  Sc,  Ti \\
 42 & Ca,  K,  Sc,  Ti \\
 43 & Ca,  Cl,  K,  Sc,  Ti \\
 44 & Ca,  K,  Sc,  Ti,  V \\
 45 & Cr,  K,  Sc,  Ti,  V \\
 46 & Cr,  K,  Sc,  Ti,  V \\
 47 & Cr,  K,  Sc,  Ti,  V \\
 48 & Cr,  K,  Sc,  Ti,  V \\
 49 & Cr,  Fe,  K,  Mn,  Sc,  Ti,  V \\
 50 & Cr,  Mn,  Sc,  Ti,  V \\
 51 & Mn,  Sc,  Ti,  V \\
 52 & Fe,  Mn,  Ti,  V \\
 53 & Cr,  Fe,  Mn,  Ti,  V \\
 54 & Cr,  Fe,  Mn,  V \\
 55 & Cr,  Fe,  Mn,  Ti,  V \\
 56 & Cr,  Fe,  Mn,  Ni,  Sc,  Ti,  V \\
 57 & Cr,  Cu,  Fe,  Mn,  Ni,  Ti,  V,  Zn \\
 58 & Cr,  Cu,  Fe,  Mn,  Ni,  Ti,  V \\
 59 & Cr,  Cu,  Fe,  Mn,  Ni,  V \\
 60 & Cr,  Cu,  Fe,  Mn,  Ni,  Ti,  V,  Zn 
\enddata
\tablenotetext{a}{All elements of the given mass for which the 
\citet{Ful82} rates were included.}
\end{deluxetable}

\clearpage

\clearpage

\clearpage

\clearpage

\clearpage

\clearpage

\clearpage

\clearpage

\clearpage

\end{document}